\documentclass[aps,prb,showpacs,twocolumn,floats]{revtex4}
\usepackage{amssymb}

\usepackage{hyperref}
\usepackage{graphicx}
\usepackage{dcolumn}
\usepackage{bm}

\input{tcilatex}

\begin{document}

\title{Dipolar Ordering and Quantum Dynamics of Domain Walls in Mn-12 Acetate}
\author{D. A. Garanin and E. M. Chudnovsky}
\affiliation{\mbox{Department of Physics and Astronomy,}
\mbox{Lehman College, City University of New York,} \\ \mbox{250
Bedford Park Boulevard West, Bronx, New York 10468-1589, U.S.A.}}
\date{\today}

\begin{abstract}
We find that dipolar interactions favor ferromagnetic ordering of elongated
crystals of Mn$_{12}$ Acetate below 0.8 K. Ordered crystals must possess
domain walls. Motion of the wall corresponds to a moving front of
Landau-Zener transitions between quantum spin levels. Structure and mobility
of the wall are computed. The effect is robust with respect to inhomogeneous
broadening and decoherence.
\end{abstract}
\maketitle


\section{Introduction}

Molecular magnets exhibit quantum dynamics at the macroscopic level. The
best-known expression of such a dynamics is the staircase magnetization
curve that one observes on changing the magnetic field. \cite
{frisartejzio96prl,heretal96epl,thoetal96nat} The steps occur due to
Landau-Zener transitions between crossing quantum spin levels. \cite
{chutej06book} It has been previously demonstrated that dipole-dipole
interactions in molecular magnets lead to ferro- or antiferromagnetic
ordering of spins at low temperature. \cite
{feralo00prb,marchuaha01epl,moretal03prl,evaetal04prl,luisetal05prl} The
Curie temperature as high as 0.9 K was reported in neutron scattering
experiments on Mn$_{12}$ Acetate.\cite{luisetal05prl} In this paper we
re-examine the effect of dipolar interactions in Mn$_{12}$ within numerical
model that treats spin-10 clusters as point magnetic dipoles located at the
sites of a body centered tetragonal lattice. We find that elongated crystals
must order ferromagnetically below 0.8 K.

It has been noticed in the past \cite{prosta98prl} that magnetic relaxation
in molecular magnets is a collective effect. Indeed, the change of the spin
state of one molecule results in the change of the long-range dipolar field
acting on other spins. When this change in the local dipolar field causes
crossing of spin levels at a certain crystal site, the spin state of the
molecule at that site changes as well. Quantum many-body Landau-Zener
dynamics of molecular magnets has been intensively studied in recent years
by means of Monte Carlo simulations \cite
{alofer01prl,feralo05prb,cucetal99epjb} and by analytical methods. \cite
{garsch05prb,voggar06prb} In this paper we employ analytical model that
takes into account both, local spin transitions and the long-range dynamics
of the dipolar field. Within such a model it becomes obvious that existing
Monte Carlo simulations of collective spin dynamics of molecular magnets
have missed an essential feature of that process: Below ordering temperature
the relaxation may occur via propagation of a domain wall (DW) separating
spin-up and spin-down regions. Unlike domain-wall motion in conventional
ferromagnets, the dynamics of the domain wall in a molecular magnet is
entirely quantum. It is driven by quantum transitions between spin levels
that are crossed in a deterministic manner in space and time by a
propagating wave of the dipolar magnetic field. Note that a propagating
front of the magnetization reversal has been recently observed in Mn$_{12}$
crystals and interpreted as magnetic deflagration. \cite
{suzetal05prl,heretal05prl, garchu07prb} The latter is a classical
phenomenon equivalent to the flame propagation, with the Zeeman energy
playing the role of the chemical energy. Quantum mechanics enters the
deflagration problem only through the reduction of the energy barrier near
the tunneling resonance. On the contrary, the phenomenon described in this
paper has quantum origin. It corresponds to a wave of Landau-Zener
transitions generated by dipole-dipole interaction between magnetic
molecules in a crystal.

We consider quantum tunneling between two nearly degenerate ground states $%
\left| \pm S\right\rangle $ of magnetic molecules at low temperatures,
interacting with each other as magnetic dipoles. The molecules are then
effectively described by spin 1/2 instead of spin 10. In the absence of
quantum tunneling between $\left| \pm S\right\rangle $ these states do not
communicate with each other so that any initial distribution of molecules in
spin-up and spin-down sates will be preserved. The measure of communication
between $\left| \pm S\right\rangle $ is their tunnel splitting $\Delta $.
The effects described in this paper, such as ferromagnetic ordering and
motion of domain walls, can be observed only if $\Delta $ is sufficiently
large. In Mn$_{12}$ it can be controlled by the transverse magnetic field.
Since we are interested in the motion of domain walls, we choose elongated
sample in the shape of a long cylinder of length $L$ and radius $R,$ the
quantization axis of spins being directed along the $z$-axis of the
cylinder. We restrict our consideration to the states only weakly nonuniform
at the lattice scale, so that spins in macroscopic regions are parallel to
each other. This can be achieved by either polarizing spins by the external
magnetic field or through ferromagnetic order which, as we shall see below,
plays an important role al low temperatures where many experiments were
performed. We further simplify the problem by ignoring inhomogeneities along
the perpendicular axes $x$ and $y,$ so that $\sigma _{z}\equiv \left\langle
S_{z}\right\rangle /S$ depends on $z$ only.

\section{The model}

\subsection{The density matrix equation}

The effective Hamiltonian of one magnetic molecule at site $i$ within the
MFA, taking into account only the two ground states $\left| \pm
S\right\rangle $ of a molecular magnet at low temperature, can be formulated
in terms of pseudospin (below spin) \ $\mathbf{\hat{\sigma}}$ as
\begin{equation}
\hat{H}_{\mathrm{eff}}=-\frac{1}{2}W\hat{\sigma}_{z}-\frac{1}{2}\Delta \hat{%
\sigma}_{x}.  \label{Ham}
\end{equation}
Here $W=2Sg\mu _{\mathrm{B}}B_{z}$ is the energy bias that generally depends
on time via the total longitudinal field $B_{z}$ including the external and
dipolar fields. $\Delta $ is tunnel splitting defined by the uniaxial
anisotropy $D$ and the terms in the Hamiltonian that cause tunneling, e.g.,
transverse field $B_{\bot }.$ Finally, $\hat{\sigma}_{z},$ $\hat{\sigma}_{x}
$ are Pauli matrices. The energy levels of this Hamiltonian for an
instantaneous value of $W$ are
\begin{equation}
\varepsilon _{\pm }=\pm \frac{1}{2}\hbar \omega _{0},\qquad \omega _{0}=%
\frac{1}{\hbar }\sqrt{W^{2}+\Delta ^{2}},  \label{epspmDef}
\end{equation}
where $\omega _{0}$ is the corresponding transition frequency.

The density-matrix equation (DME) for the spin in the time-dependent
adiabatic basis, formed by the instantaneous eigenstates $\left| \chi _{\pm
}\right\rangle $ of $\hat{H}_{\mathrm{eff}}$, has the form
\begin{eqnarray}
&&\frac{d}{dt}\rho _{++}=\left( \langle \dot{\chi}_{+}\left| \chi
_{+}\right\rangle +\langle \chi _{+}\left| \dot{\chi}_{+}\right\rangle
\right) \rho _{++}+  \nonumber \\
&&\langle \dot{\chi}_{+}\left| \chi _{-}\right\rangle \rho _{-+}+\rho
_{+-}\langle \chi _{-}\left| \dot{\chi}_{+}\right\rangle -\Gamma _{-+}\rho
_{++}+\Gamma _{+-}\rho _{--}  \nonumber \\
&&\frac{d}{dt}\rho _{+-}=\left( \langle \dot{\chi}_{+}\left| \chi
_{+}\right\rangle +\langle \chi _{-}\left| \dot{\chi}_{-}\right\rangle
\right) \rho _{+-}+\rho _{++}\langle \chi _{+}\left| \dot{\chi}%
_{-}\right\rangle  \nonumber \\
&&+\langle \dot{\chi}_{+}\left| \chi _{-}\right\rangle \rho _{--}-\left[
i\omega _{0}+\frac{1}{2}\left( \Gamma _{-+}+\Gamma _{+-}\right) \right] \rho
_{+-},  \label{DME-GGResAdi0}
\end{eqnarray}
where $\Gamma _{-+}$,$\Gamma _{+-}$ are up and down relaxation rates for the
levels $\varepsilon _{\pm },$ satisfying the detailed balance condition $%
\Gamma _{+-}=e^{-\hbar \omega _{0}/(k_{\mathrm{B}}T)}\Gamma _{-+}.$ The
elements of the density matrix satisfy $\rho _{++}+\rho _{--}=1$ and $\rho
_{-+}=\left( \rho _{+-}\right) ^{\ast }.$

Taking time derivative of the spin expectation value $\mathbf{\sigma }=%
\mathrm{Tr}(\rho \hat{\mathbf{\sigma }})$ and using the relation
\begin{equation}
\mathbf{\sigma =}\left( \rho _{+-}+\rho _{-+}\right) \mathbf{e}_{x}+i\left(
\rho _{+-}-\rho _{-+}\right) \mathbf{e}_{y}+\left( \rho _{--}-\rho
_{++}\right) \mathbf{e}_{z},
\end{equation}
one finds that in the chosen time-dependent frame the DME describes damped
precession of $\mathbf{\sigma }$ about the effective field $\mathbf{\omega }%
_{0}+\dot{\theta}\mathbf{e}_{y}$, where
\begin{equation}
\mathbf{\omega }_{0}=\frac{1}{\hbar }\left( \Delta \mathbf{e}_{x}+W\mathbf{e}%
_{z}\right)  \label{omegaoiDef}
\end{equation}
and $\cos \theta =W/\sqrt{W^{2}+\Delta ^{2}}$ describes the orientation of $%
\mathbf{\omega }_{0}.$ Switching to the laboratory coordinate frame (which
amounts to dropping $\ $non-adiabatic term $\dot{\theta}$ in the effective
field) one obtains
\begin{eqnarray}
\mathbf{\dot{\sigma}} &=&\left[ \mathbf{\sigma }\times \mathbf{\omega }_{0}%
\right]  \nonumber \\
&-&\frac{\Gamma }{2}\left( \mathbf{\sigma }-\frac{\mathbf{\omega }_{0}\cdot
\mathbf{\sigma }}{\omega _{0}^{2}}\mathbf{\omega }_{0}\right) -\Gamma \left(
\frac{\mathbf{\omega }_{0}\cdot \mathbf{\sigma }}{\omega _{0}^{2}}\mathbf{%
\omega }_{0}-\mathbf{\sigma }_{0}\right) ,  \label{DME-sigmaEqVec}
\end{eqnarray}
where $\Gamma =\Gamma _{-+}+$ $\Gamma _{+-}$ and $\mathbf{\sigma }_{0}$ is
the thermal equilibrium value of the pseudospin, corresponding to the
instantaneous value and direction of $\mathbf{\omega }_{0}.$ The second term
in this equation corresponds to the relaxation of the spin component
perpendicular to $\mathbf{\omega }_{0}$ while the third term corresponds to
relaxation along $\mathbf{\omega }_{0},$ the latter being twice as fast as
the former.

The equilibrium solution of Eq. (\ref{DME-sigmaEqVec}) has the form
\begin{equation}
\mathbf{\sigma }=\mathbf{\sigma }_{0}=\sigma _{0}\frac{\mathbf{\omega }_{0}}{%
\omega _{0}},\qquad \sigma _{0}=\tanh \frac{\hbar \omega _{0}}{2k_{\mathrm{B}%
}T}.  \label{sigmaiEquil}
\end{equation}
Remember that the equations above are written for the spin on a site $i,$
although the index $i$ is not explicitly written for brevity. The equation
of motion (\ref{DME-sigmaEqVec}) for spin $i$ couples to those for all other
spins $j$ via the magnetostatic contribution in $W$ considered in the next
section.

In transverse field $B_{\bot }$ satisfying $\hbar \omega _{0}\ll g\mu
_{B}B_{\bot }$ the relaxation rate $\Gamma $ due to direct processes is
given by

\begin{equation}
\Gamma =\frac{S^{2}\Delta ^{2}\omega _{0}\left( g\mu _{B}B_{\bot }\right)
^{2}}{12\pi E_{t}^{4}}\coth \frac{\hbar \omega _{0}}{2k_{\mathrm{B}}T},
\label{Gammai}
\end{equation}
where $E_{t}\equiv \left( \rho v_{t}^{5}\hbar ^{3}\right) ^{1/4}$ is a
characteristic energy. For Mn$_{12}$ $E_{t}/k_{\mathrm{B}}\simeq 150$ K but
the factor 1/2 should be introduced in $\Gamma $ since there is only one
transverse sound mode. Usually direct rates are proportional to $\omega
_{0}^{3}$ but in the case of tunneling there is an additional factor $\omega
_{0}^{-2}$ accounting for the decrease of hybridization of the two levels
with increasing the energy bias $W.$

The model formulated above describes the Landau-Zener effect in the case of
time-dependent energy bias $W$ that crosses zero, as well as relaxation. In
the presence of dipolar coupling this model describes magnetic ordering \
and domain-wall dynamics. Especially interesting is possible interplay
between the DW dynamics and LZ effect that can possibly lead to Landau-Zener
fronts. The model can be extended by including heating the sample as the
result of spin relaxation. This leads to formation of deflagration fronts if
the relaxation rate strongly increases with temperature.

\subsection{The dipolar field}

The energy bias $W$ in the equations above at the site $i$ is given by
\begin{equation}
W_{i}=2Sg\mu _{\mathrm{B}}\left( B_{z}+B_{i,z}^{(\mathrm{D})}\right) \equiv
W_{\mathrm{ext}}+W_{i}^{(\mathrm{D})},  \label{WiDef}
\end{equation}
where $B_{z}$ is the $z$ component of the external field and $B_{i,z}^{(%
\mathrm{D})}$ is the dipolar field at site $i.$ The dipolar component of the
bias is given by
\begin{equation}
W_{i}^{(\mathrm{D})}=2E_{\mathrm{D}}D_{i,zz},\qquad D_{i,zz}\equiv
\sum_{j}\phi _{ij}\sigma _{jz},  \label{WiDDef}
\end{equation}
where $E_{\mathrm{D}}\equiv \left( g\mu _{\mathrm{B}}S\right) ^{2}/v_{0}$ is
the dipolar energy, $v_{0}$ is the unit-cell volume, and
\begin{equation}
\phi _{ij}=v_{0}\frac{3\left( \mathbf{e}_{z}\cdot \mathbf{n}_{ij}\right)
^{2}-1}{r_{ij}^{3}},\qquad \mathbf{n}_{ij}\equiv \frac{\mathbf{r}_{ij}}{%
r_{ij}}  \label{phizijDef}
\end{equation}
is the dimensionless dipole-dipole interaction between the spins at sites $i$
and $j\neq i.$ The $z$ component of the dipolar field itself is given by
\begin{equation}
B_{i,z}^{(\mathrm{D})}=\frac{Sg\mu _{\mathrm{B}}}{v_{0}}D_{i,zz}.
\label{BDiz}
\end{equation}

To calculate the dipolar field, one can introduce a macroscopic sphere of
radius $r_{0}$ satisfying $v_{0}^{1/3}\ll r_{0}\ll L$ around the site $i,$
where $L$ is the (macrocopic) linear size of the sample. The field from the
spins at sites $j$ inside this sphere can be calculated by direct summation
over the lattice, whereas the field from the spins outside the sphere can be
obtained by integration. The details are given in the Appendix. In
particular, for a uniformly magnetized ellipsoid the total result has the
form
\begin{equation}
D_{zz}\equiv \sigma _{z}\sum_{j}\phi _{ij}=\bar{D}_{zz}\sigma _{z},
\label{DzzEllipsoid}
\end{equation}
independently of $i,$ where
\begin{equation}
\bar{D}_{zz}=\bar{D}_{zz}^{(\mathrm{sph})}+4\pi \nu \left(
1/3-n^{(z)}\right)   \label{DzzEllipsoid1}
\end{equation}
and $\nu $ is the number of molecules per unit cell. For the demagnetizing
factor one has $n^{(z)}=0,$ $1/3,$ and 1 for a cylinder, sphere, and disc,
respectively. One obtains $\bar{D}_{zz}^{(\mathrm{sph})}=0$ for a simple
cubic lattice, $\bar{D}_{zz}^{(\mathrm{sph})}<0$ for a tetragonal lattice
with $a=b>c$, and $\bar{D}_{zz}^{(\mathrm{sph})}>0$ for $a=b<c.$
\begin{equation}
E_{0}=-(1/2)\bar{D}_{zz}E_{D}  \label{E0Def}
\end{equation}
is the dipolar energy per site.

The two best known molecular magnets are Mn$_{12}$ and Fe$_{8},$ both having
total spin $S=10$. Mn$_{12}$ crystallizes in a body-centered tetragonal
(bct) lattice with $a=b=17.319$ \AA ,\ $c=12.388$ \AA\ ($c$ being the easy
axis) and the unit-cell volume $v_{0}=abc=3716$ \AA $^{3}$, with two
molecules per unit cell, $\nu =2$. Fe$_{8}$ has a triclinic lattice with $%
a=10.52$ \AA\ ($a$ being the easy axis), $b=14.05$ \AA , $c=15.00$ \AA , $%
\alpha =89.9%
{{}^\circ}%
,$ $\beta =109.6%
{{}^\circ}%
,$ $\gamma =109.3%
{{}^\circ}%
$ and $v_{0}=abc\sin \alpha \sin \beta \sin \gamma =1971$ \AA $^{3}$. The
characteristic dipolar energies thus are $E_{\mathrm{D}}/k_{\mathrm{B}%
}=0.0671$ K for Mn$_{12}$ and $E_{\mathrm{D}}/k_{\mathrm{B}}=0.126$ K for Fe$%
_{8}.$

For Fe$_{8}$ direct numerical calculation yields \cite{garsch05prb} $\bar{D}%
_{zz}^{(\mathrm{sph})}=4.072,$ thus for the cylinder Eq. (\ref{DzzEllipsoid1}%
) yields $\bar{D}_{zz}^{(\mathrm{cyl})}=8.261.$ Our result $%
E_{0}=-4.131E_{D} $ for the elongated Fe$_{8}$ crystal is in qualitative
accord with $E_{0}=-4.10E_{D}$ of Ref.\ \onlinecite{marchuaha01epl}.

For Mn$_{12}$ one obtains $\bar{D}_{zz}^{(\mathrm{sph})}=2.155$ that results
in $\bar{D}_{zz}^{(\mathrm{cyl})}=10.53$ for a cylinder. Then Eq. (\ref{BDiz}%
) yields the dipolar field $B_{z}^{(\mathrm{D})}\simeq 0.0526$ T in an
elongated sample. On the top of it, there is a weak ferromagnetic exchange
interaction between the neighboring Mn$_{12}$ molecules that creates an
effective field 7 G from each neighbor. \cite{parketal02prb} With 8 nearest
neighbors in the bct lattice, this effectively adds 1.12 to $\bar{D}_{zz}$
in the ferromagnetic state. Thus for Mn$_{12}$ one obtains effectively $\bar{%
D}_{zz}^{(\mathrm{cyl})}\simeq 11.65.$

It can be shown that all other types of ordering have a lower value of $\bar{%
D}_{zz}$ and thus a higher value of the ground-state energy $E_{0}$ for both
Mn$_{12}$ and Fe$_{8}$ of a cylindric shape. For Fe$_{8}$ the state with
ferromagnetically ordered planes alternating in the $c$ direction has $\bar{D%
}_{zz}=8.18,$ the dipolar field being shape independent. This value is very
close to $8.261$ for the ferromagnetically ordered cylinder and it has only
slightly higher energy $E_{0}.$ The states with ferromagnetic planes
alternating in the $b$ direction has $\bar{D}_{zz}=8.12,$ while the state of
ferromagnetic chains directed along the $a$ axis and alternating in $b$ and $%
c$ direction has $\bar{D}_{zz}=8.01.$ One can see that a crystal of Fe$_{8}$
cooled in zero field will show a random mixture of different types of
ordering. On the other hand, as different kinds of ordered states are
separated by energy barriers, prepared ordered states should be robust
metastable states. In particular, ferromagnetically ordered state of an
elongated Fe$_{8}$ crystal, obtained by cooling in the magtic field, should
be stable after removing this field at low temperatures.

For Mn$_{12},$ states with ferromagnetically ordered planes alternating in
the $a$ or $b$ directions in \emph{each} sublattice have $\bar{D}%
_{zz}=9.480, $ independently of the shape and of the exchange interaction.
The state with alternating chains directed along the $c$ direction has a
very close value $\bar{D}_{zz}=9.475.$ For the two-sublattice
antiferromagnetic ordering one obtains $\bar{D}_{zz}=8.102.$ All these
values are essentially lower than the dipolar field $\bar{D}_{zz}^{(\mathrm{%
cyl})}\simeq 11.65$ for a cylinder. Thus one can expect that elongated
crystals of Mn$_{12}$ will order ferromagnetically without competition of
other states.

On the other hand, for a spherical shape the states with alternating
ferromagnetically ordered planes and chains are energetically more favorable
for both Fe$_{8}$ and Mn$_{12}.$

For a cylinder magnetized with $\sigma _{z}=$ $\sigma _{z}(z),$ the field
along the symmetry axis has the form
\begin{equation}
D_{zz}(z)=\nu \int_{-L/2}^{L/2}dz^{\prime }\frac{2\pi R^{2}\sigma
_{z}(z^{\prime })}{\left[ \left( z^{\prime }-z\right) ^{2}+R^{2}\right]
^{3/2}}-k\sigma _{z}(z),  \label{DzzCylinder}
\end{equation}
where
\begin{equation}
k\equiv 8\pi \nu /3-\bar{D}_{zz}^{(\mathrm{sph})}=4\pi \nu -\bar{D}_{zz}^{(%
\mathrm{cyl})}>0,  \label{kDef}
\end{equation}
$k=14.6$ for Mn$_{12}$ and $k=4.31$ for Fe$_{8}.$ For a uniformly polarized
cylinder Eq. (\ref{DzzCylinder}) yields
\begin{eqnarray}
D_{zz}(z) &=&2\pi \nu \left( \frac{z+L/2}{\sqrt{\left( z+L/2\right)
^{2}+R^{2}}}\right.  \nonumber \\
&&\qquad -\left. \frac{z-L/2}{\sqrt{\left( z-L/2\right) ^{2}+R^{2}}}\right)
\sigma _{z}-k\sigma _{z}.  \label{DzzCylinderUniform}
\end{eqnarray}
In the depth of a long cylinder, $L\gg R,$ the dipolar field is $%
D_{zz}=\left( \bar{D}_{zz}^{(\mathrm{sph})}+4\pi \nu /3\right) \sigma _{z},$
in accordance with Eq. (\ref{DzzEllipsoid1})$.$

At an end of a long cylinder one has $D_{zz}=\left( \bar{D}_{zz}^{(\mathrm{%
sph})}-2\pi \nu /3\right) \sigma _{z}$ that can have the sign opposite to
that in the depth of the sample. In particular, for Fe$_{8}$ with $\bar{D}%
_{zz}^{(\mathrm{sph})}=4.072$ and $\nu =1$ one has $D_{zz}=1.98\sigma _{z}.$
Thus a homogeneously magnetized state in zero field is stable. To the
contrary, for Mn$_{12}$ with $\bar{D}_{zz}^{(\mathrm{sph})}=2.155$ and $\nu
=2$ one has $D_{zz}=-2.03\sigma _{z}$. This means that a homogeneously
magnetized state in zero external field is unstable with respect to domain
formation, beginning in the vicinity of the ends of the cylinder. At some
point near the end of the crystal the resonance condition $D_{zz}(z)=0$ is
satisfied that leads to spin tunneling and the decay of the initially
homogeneously magnetized state.

If there is a domain wall at $z=0$ ($\sigma _{z}\rightarrow \sigma _{\infty
} $ for $z\rightarrow -\infty $) in a long cylinder that is narrow in
comparizon to $R,$ Eq. (\ref{DzzCylinder}) far from the ends yields
\begin{equation}
D_{zz}(z)=-4\pi \nu \frac{z-z_{0}}{\sqrt{\left( z-z_{0}\right) ^{2}+R^{2}}}%
\sigma _{\infty }-k\sigma _{z}(z).  \label{DzzCylinderNarrowWall}
\end{equation}
Since the coefficient of the local term is negative, the latter changes in
the opposite direction with respect to the first term. This creates three
zeros of $D_{zz}(z).$ However, such exotic DWs do not exist, as we will see
below. Thermodynamically stable domain walls have only one zero of $%
D_{zz}(z),$ thus their width at low temperatures is of order $R.$

\section{Dipolar ordering and static domain wall profile}

\subsection{Ferromagnetic ordering}

Dipolar field causes ferromagnetic ordering in elongated crystals of
molecular magnets that is described within the MFA by the Curie-Weiss
equation following from Eqs. (\ref{sigmaiEquil}), (\ref{WiDef}), and (\ref
{DzzEllipsoid}). This equation for a homogeneously magnetized sample of an
ellipsoidal shape has the form
\begin{equation}
\sigma _{z}=\frac{W_{\mathrm{ext}}+2E_{\mathrm{D}}\bar{D}_{zz}\sigma _{z}}{%
\hbar \omega _{0}}\tanh \frac{\hbar \omega _{0}}{2k_{\mathrm{B}}T},
\label{CWEq}
\end{equation}
where $\hbar \omega _{0}=\sqrt{\left( W_{\mathrm{ext}}+2E_{\mathrm{D}}\bar{D}%
_{zz}\sigma _{z}\right) ^{2}+\Delta ^{2}.}$ This equation is similar to that
for the Ising model in a transverse field, here $\Delta $. The Curie
temperature satisfies the equation
\begin{equation}
1=\frac{2E_{\mathrm{D}}\bar{D}_{zz}}{\Delta }\tanh \frac{\Delta }{2k_{%
\mathrm{B}}T_{\mathrm{C}}}.  \label{TCEq}
\end{equation}
This equation has a solution only for $\Delta <E_{\mathrm{D}}\bar{D}_{zz}$
as the transverse field tends to suppress the phase transition. In the
actual case $\Delta \ll E_{\mathrm{D}}\bar{D}_{zz}$ one obtains
\begin{equation}
T_{\mathrm{C}}=E_{\mathrm{D}}\bar{D}_{zz}/k_{\mathrm{B}}.
\label{TCsmallDelta}
\end{equation}
For a Mn$_{12}$ one has $E_{\mathrm{D}}/k_{\mathrm{B}}\simeq 0.0671$ K. Thus
for a cylinder ($\bar{D}_{zz}\simeq 11.65$) Eq. (\ref{TCsmallDelta}) yields $%
T_{\mathrm{C}}\simeq 0.782$ K. This is close to the value 0.9 K reported in
Ref. \onlinecite{luisetal05prl}. Note that the MFA does not account for
fluctuations that usually lower $T_{\mathrm{C}}.$ However, as we have seen
above, for the Mn$_{12}$ cylinder the main contribution to the dipolar field
comes from a great number of distant spins, thus the MFA should work well.
The small disagreement in $T_{\mathrm{C}}$ can be attributed to
approximating magnetic molecules by point dipoles in our calculations.

The equilibrium value of $\sigma _{z}$ in zero field below $T_{\mathrm{C}}$
satisfies the equation
\begin{equation}
1=\frac{2E_{\mathrm{D}}\bar{D}_{zz}}{\hbar \omega _{0}}\tanh \frac{\hbar
\omega _{0}}{2k_{\mathrm{B}}T},  \label{CWEqordered}
\end{equation}
where $\hbar \omega _{0}=\sqrt{\left( 2E_{\mathrm{D}}\bar{D}_{zz}\sigma
_{z}\right) ^{2}+\Delta ^{2}}.$ In the realistic case $\Delta \ll E_{\mathrm{%
D}}\bar{D}_{zz}$ not too close to the Curie point one can neglect $\Delta
^{2}$ in $\hbar \omega _{0}$, then Eq. (\ref{CWEqordered}) defining the spin
polarization below $T_{\mathrm{C}}$ takes the form
\begin{equation}
\sigma _{z}=\tanh \frac{E_{\mathrm{D}}\bar{D}_{zz}\sigma _{z}}{k_{\mathrm{B}%
}T}.
\end{equation}
For Mn$_{12}$ the ratio $\Delta /\left( 2E_{\mathrm{D}}\bar{D}_{zz}\right) $
is small. Even for $B_{\perp }=5$ T that corresponds to $h_{\perp }\equiv
g\mu _{\mathrm{B}}B_{\perp }/(2SD)=0.61,$ one has $\Delta /k_{\mathrm{B}}=$%
0.00166 K and thus $\Delta /\left( 2E_{\mathrm{D}}\bar{D}_{zz}\right) \simeq
2.5\times 10^{-3}.$

\subsection{Static domain wall profile}

The magnetization profile $\sigma _{z}(z)$ in a domain wall joining regions
with $\sigma _{z}=\pm $ $\sigma _{\infty }=\pm \sigma _{z}^{\mathrm{eq}}$ in
zero field satisfies the equation
\begin{equation}
\sigma _{z}=\frac{2E_{\mathrm{D}}D_{zz}(z)}{\hbar \omega _{0}}\tanh \frac{%
\hbar \omega _{0}}{2k_{\mathrm{B}}T},  \label{DWProfileEq}
\end{equation}
where $\hbar \omega _{0}=\sqrt{\left[ 2E_{\mathrm{D}}D_{zz}(z)\right]
^{2}+\Delta ^{2}}.$ Here $D_{zz}(z)$ is given by Eq. (\ref{DzzCylinder})
that makes Eq. (\ref{DWProfileEq}) an integral equation. By symmetry, $%
D_{zz}(z)$ goes through zero in the center of the DW. At very low
temperatures, $T\ll \Delta /k_{\mathrm{B}},$ the argument of tanh is always
large, thus Eq. (\ref{DWProfileEq}) simplifies to
\begin{equation}
\sigma _{z}=\frac{2E_{\mathrm{D}}D_{zz}(z)}{\sqrt{\left[ 2E_{\mathrm{D}%
}D_{zz}(z)\right] ^{2}+\Delta ^{2}}}.  \label{DWProfileEqLT}
\end{equation}
In this limit the spin length is nearly constant, $\sigma \cong 1,$ so that $%
\sigma _{x}\cong \sqrt{1-\sigma _{z}^{2}}.$ The spin-polarization deficit in
the wall is defined by
\begin{eqnarray}
1-\sigma &=&1-\tanh \frac{\sqrt{\left[ 2E_{\mathrm{D}}D_{zz}(z)\right]
^{2}+\Delta ^{2}}}{2k_{\mathrm{B}}T}  \nonumber \\
\qquad \qquad &\cong &\frac{2k_{\mathrm{B}}T}{\sqrt{\left[ 2E_{\mathrm{D}%
}D_{zz}(z)\right] ^{2}+\Delta ^{2}}}\ll 1  \label{MagDeficit}
\end{eqnarray}
and it reaches the maximum at the DW center. With the help of Eq. (\ref
{DWProfileEqLT}) one finds
\begin{equation}
1-\sigma \cong \frac{2k_{\mathrm{B}}T}{\Delta }\sqrt{1-\sigma _{z}^{2}}.
\label{MagDefsig}
\end{equation}

In the opposite limit $\Delta /k_{\mathrm{B}}\ll T$ Eq. (\ref{DWProfileEq})
simplifies to
\begin{equation}
\sigma _{z}=\tanh \frac{E_{\mathrm{D}}D_{zz}(z)}{k_{\mathrm{B}}T}.
\label{DWProfileEqHT}
\end{equation}
The transverse spin component in this case is given by
\begin{equation}
\sigma _{x}=\frac{\Delta }{2E_{\mathrm{D}}D_{zz}(z)}\tanh \frac{E_{\mathrm{D}%
}D_{zz}(z)}{k_{\mathrm{B}}T},  \label{DWProfileEqHTx}
\end{equation}
and $\sigma _{x}=\Delta /(2k_{\mathrm{B}}T)\ll 1$ at the center of the DW.
For Mn$_{12}$ $\sigma _{x}$ is very small except of very low temperatures
and very large transverse fields. Such a domain wall is the linear
(Ising-like) domain wall found in ferromagnets in a narrow temperature range
below $T_{\mathrm{C}},$ \cite
{bulgin64jetp,lajnie79,koegarharjah93prl,harkoegar95prb} as well as in the
low-temperature strong-anisotropy ferromagnet GdCl$_{3}.$ \cite{grakoe89}
Absence of a strong short-range exchange interaction in molecular magnets
makes domain walls linear practically in the whole range below $T_{\mathrm{C}%
}.$

\begin{figure}[t]
\includegraphics[angle=-90,width=8cm]{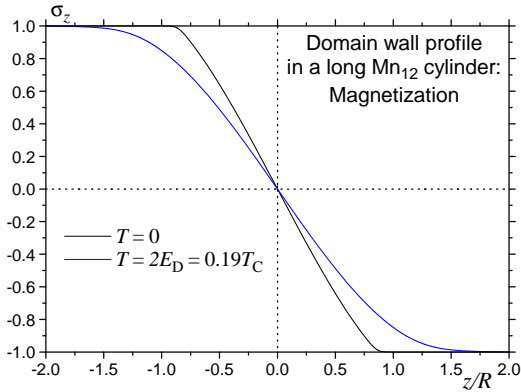}
\caption{{}Magnetization profile of a domain-wall in a Mn$_{12}$ cylinder at
two different temperatures.}
\label{Fig-DW_profile_Temperatures}
\end{figure}
\begin{figure}[t]
\includegraphics[angle=-90,width=8cm]{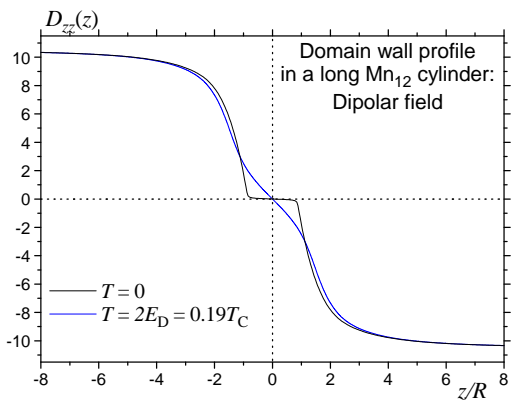}
\caption{{}Dipolar-field profile of a domain-wall in a Mn$_{12}$ cylinder at
two different temperatures.}
\label{Fig-DW-Dzz_profile_Temperatures}
\end{figure}

For Mn$_{12}$ the splitting $\Delta $ is typically so small that one can
consider the limit $T\rightarrow 0$ of Eq. (\ref{DWProfileEqHT}). Then three
regions arise: (i) left from the DW where $\sigma _{z}=1$ and the argument
of tanh is infinite, that is, $D_{zz}(z)>0$; (ii) inside the DW where $%
\left| \sigma _{z}\right| <1$ and thus $D_{zz}(z)=0;$ right from the DW
where $\sigma _{z}=-1$ and $D_{zz}(z)<0$. Using Eq. (\ref{DzzCylinder}) with
$L=\infty $ one obtains the integral equation for the nontrivial region
inside the DW centered at $z=0$:
\begin{eqnarray}
&&0=\nu \int_{-l}^{l}dz^{\prime }\frac{2\pi R^{2}\sigma _{z}(z^{\prime })}{%
\left[ \left( z^{\prime }-z\right) ^{2}+R^{2}\right] ^{3/2}}-k\sigma _{z}(z)
\nonumber \\
&&-2\pi \nu \left( \frac{z-l}{\sqrt{(z-l)^{2}+R^{2}}}+\frac{z+l}{\sqrt{%
(z+l)^{2}+R^{2}}}\right) ,  \label{sigmazEqT0}
\end{eqnarray}
where $l$ is the DW width that should follow from the equation above and the
boundary conditions are $\sigma _{z}(\pm l)=\mp 1$. One can see that still
in for $T\rightarrow 0$ the DW profile results from an essentially integral
equation that does not have a general analytical solution, although
analytical solutions in limiting cases do exist. Rewriting Eq. (\ref
{sigmazEqT0}) in terms of $\tilde{z}\equiv z/R,$ one can show that the DW
profiles scales with the cylinder radius $R,$ i.e., $l\sim R.$

Numerical solution using relaxation to the equilibrium described by Eq. (\ref
{DWProfileEq}) with proper boundary conditions at the ends of a long Mn$%
_{12} $ cylinder is shown in Figs. \ref{Fig-DW_profile_Temperatures} and \ref
{Fig-DW-Dzz_profile_Temperatures}. One can see that at $T=0$ the
magnetization profile turns to $\pm 1$ beyond the region of the DW of the
length $l\sim R$, in accordance with Eq. (\ref{sigmazEqT0}). At finite
temperatures the solution for $\sigma _{z}(z)$ is more resembling a $\tanh $%
. The DW width $l$ increases with temperature and diverges at $T_{\mathrm{C}%
}.$ The solution for the dipolar field in Fig. \ref
{Fig-DW-Dzz_profile_Temperatures} shows that $D_{zz}(z)\rightarrow 0$ at $%
T\rightarrow 0$ inside the domain wall while it is approaching the
asymptotic value $\pm 11.65$ in the domains. At finite temperatures the
dependence $D_{zz}(z)$ smoothens out. Overall the region of inhomoheneity of
the dipolar field is broader than that of the magnetization.

\begin{figure}[t]
\includegraphics[angle=-90,width=8cm]{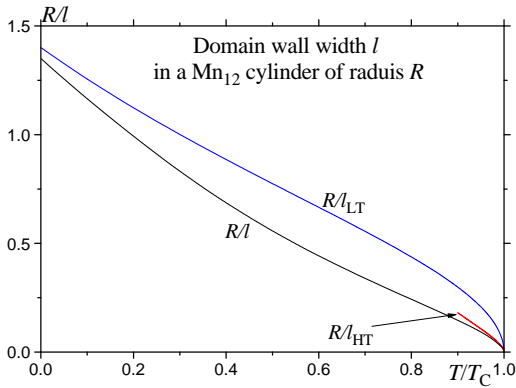}
\caption{{}Temperature dependence of the DW width $l$ and its low- and
high-temperature forms in a Mn$_{12}$ cylinder.}
\label{Fig-l}
\end{figure}

The DW width $l$ can be defined as a slope
\begin{equation}
l^{-1}=\frac{1}{\sigma _{\infty }}\left. \frac{d\sigma _{z}}{dz}\right|
_{z=0}.
\end{equation}
Some results for $l$ can be obtained analytically. At $T\ll T_{\mathrm{C}},$
assuming that the magnetization profile in the DW is close to the piece-wise
linear, one obtains from Eq.\ (\ref{DWProfileEqHT})
\begin{eqnarray}
\frac{l_{\mathrm{LT}}}{R} &=&\left[ \left( 4\pi \nu \right) ^{2}\left( \frac{%
k_{\mathrm{B}}T}{E_{\mathrm{D}}}+k\right) ^{-2}-1\right] ^{-1/2}  \nonumber
\\
&=&\frac{4\pi \nu E_{\mathrm{D}}+T-T_{C}}{\sqrt{\left( T_{C}-T\right) \left(
8\pi \nu E_{\mathrm{D}}+T-T_{C}\right) }},  \label{l}
\end{eqnarray}
in a qualitative agreement with numerical results in Fig. \ref{Fig-l}. At $%
T\rightarrow T_{\mathrm{C}}$ the solution of Eq.\ (\ref{DWProfileEqHT}) is
\begin{equation}
\sigma _{z}(z)=\sigma _{\infty }\tanh \frac{z}{l_{\mathrm{HT}}},
\end{equation}
where $\sigma _{\infty }=\sqrt{3}(T_{\mathrm{C}}/T-1)^{1/2}$ and $l_{\mathrm{%
HT}}$ satisfies
\begin{equation}
l_{\mathrm{HT}}^{2}=l_{\mathrm{LT}}^{2}\ln \frac{l_{\mathrm{HT}}^{2}}{2R^{2}}
\end{equation}
with $l_{\mathrm{LT}}$ of Eq.\ (\ref{l}) diverging as
\begin{equation}
\frac{l_{\mathrm{LT}}}{R}=\sqrt{\frac{2\pi E_{\mathrm{D}}}{k_{\mathrm{B}}(T_{%
\mathrm{C}}-T)}}.  \label{crit-l}
\end{equation}
Temperature dependence of the approximations $l_{\mathrm{LT}}$ and $l_{%
\mathrm{HT}}$, together with numerical result for the domain wall width $l$,
is shown in Fig. \ref{Fig-l}.

\section{Domain-wall dynamics}

In the sequel we will be mostly interested in the motion of domain walls
with a small speed, induced by a small external bias $B_{z}.$ In this case
the system does not deviate much from the equilibrium. In this case domain
walls in a sufficiently long sample are stationarily moving with a speed $v_{%
\mathrm{DW}}$ that is proportional to the bias field. This relation has the
form
\begin{equation}
v_{\mathrm{DW}}=\mu _{\mathrm{DW}}B_{z},  \label{muDWDef}
\end{equation}
where $\mu _{\mathrm{DW}}$ is the \emph{linear mobility} of the DW. At high
values of $B_{z}$ the dependence $v_{\mathrm{DW}}(H_{z})$ becomes nonlinear.
The linear mobility $\mu _{\mathrm{DW}}$ can be calculated using the \emph{%
static} DW profile discussed above with the help of the energy balance
argument. To this end, the LLB equation (\ref{DME-sigmaEqVec}) should be
transformed into a special form near the equilibrium.

\subsection{Landau-Lifshitz-Bloch equation near equilibrium}

One can project relaxation terms in Eq. (\ref{DME-sigmaEqVec}) on the
directions parallel and perpendicular to $\mathbf{\sigma }$ using
\begin{equation}
\mathbf{R}=\frac{\left( \mathbf{\sigma \cdot R}\right) \mathbf{\sigma }}{%
\sigma ^{2}}-\frac{\left[ \mathbf{\sigma \times }\left[ \mathbf{\sigma
\times R}\right] \right] }{\sigma ^{2}}.
\end{equation}
This results in a Landau-Lifshitz-Bloch (LLB) equation
\begin{eqnarray}
\mathbf{\dot{\sigma}} &=&\left[ \mathbf{\sigma \times \omega }_{0}\right]
\nonumber \\
&&+\Gamma \left[ \frac{\mathbf{\omega }_{0}\cdot \mathbf{\sigma }}{\omega
_{0}\sigma }\left( \frac{\sigma _{0}}{\sigma }-1\right) -\frac{1}{2}\left( 1-%
\frac{\mathbf{\omega }_{0}\cdot \mathbf{\sigma }}{\omega _{0}\sigma }\right)
^{2}\right] \mathbf{\sigma }  \nonumber \\
&&\mathbf{-}\frac{\Gamma }{2}\left( 2\sigma _{0}-\frac{\mathbf{\omega }%
_{0}\cdot \mathbf{\sigma }}{\omega _{0}}\right) \frac{\left[ \mathbf{\sigma
\times }\left[ \mathbf{\sigma \times \omega }_{0}\right] \right] }{\omega
_{0}\sigma ^{2}}.
\end{eqnarray}
Here, in contrast to ferromagnets where $\sigma \cong 1$ is enforced by a
strong exchange, $\sigma $ can essentially deviate from 1 because of the
first relaxation term. Close to equilibrium vectors $\mathbf{\omega }_{0}$
and $\mathbf{\sigma }$ are nearly collinear, thus $\mathbf{\omega }_{0}\cdot
\mathbf{\sigma }\cong \omega _{0}\sigma $ holds up to quadratic terms in
small deviations, whereas $\sigma $ is close to $\sigma _{0}.$ Thus the
equation above simplifies to
\begin{equation}
\mathbf{\dot{\sigma}}=\left[ \mathbf{\sigma \times \omega }_{0}\right]
+\Gamma \left( \frac{\sigma _{0}}{\sigma }-1\right) \mathbf{\sigma -}\frac{%
\Gamma }{2}\frac{\left[ \mathbf{\sigma \times }\left[ \mathbf{\sigma \times
\omega }_{0}\right] \right] }{\sigma \omega _{0}}.  \label{LLB}
\end{equation}
Note that $\sigma _{0}$ is formally defined by Eq. (\ref{sigmaiEquil}) and
it is a function of $\omega _{0}$ that can depend on time. Thus, in general,
$\sigma _{0}$ corresponds not to a true equilibrium but to an instantaneous
equilibrium to which the system tends. If the state of the system, as well
as the instantaneous equilibrium, is close to a true equilibrium, one can
approximately express $\mathbf{\omega }_{0}$ through $\mathbf{\sigma }$ and
small $\mathbf{\dot{\sigma}}$ that has an important application. First,
multiplying Eq. (\ref{LLB}) by $\mathbf{\sigma }$ one obtains the relation
\begin{equation}
\sigma _{0}=\sigma +\frac{1}{\Gamma }\frac{\left( \mathbf{\sigma \cdot \dot{%
\sigma}}\right) }{\sigma }.  \label{sigma0viasigma}
\end{equation}
Second, using $\mathbf{\omega }_{0}\cdot \mathbf{\sigma }\cong \omega
_{0}\sigma $ one can rewrite Eq. (\ref{LLB}) in the form
\begin{equation}
\mathbf{\dot{\sigma}}=\left[ \mathbf{\sigma \times \omega }_{0}\right]
+\Gamma \left( \frac{\sigma _{0}}{\sigma }-1\right) \mathbf{\sigma -}\frac{%
\Gamma }{2}\mathbf{\sigma }+\frac{\Gamma }{2}\frac{\sigma }{\omega _{0}}%
\mathbf{\omega }_{0}.
\end{equation}
From the latter follows
\begin{equation}
\left[ \mathbf{\sigma \times \dot{\sigma}}\right] =\mathbf{\sigma }\omega
_{0}\sigma -\mathbf{\omega }_{0}\sigma ^{2}+\frac{\Gamma }{2}\frac{\sigma }{%
\omega _{0}}\left[ \mathbf{\sigma \times \omega }_{0}\right] .
\end{equation}
Eliminating $\left[ \mathbf{\sigma \times \omega }_{0}\right] $ and using
Eq. (\ref{sigma0viasigma}), one obtains
\begin{equation}
\mathbf{\omega }_{0}=\omega _{0}\frac{\mathbf{\sigma }}{\sigma }-\frac{%
\omega _{0}^{2}}{\omega _{0}^{2}+\Gamma ^{2}/4}\frac{\left[ \mathbf{\sigma
\times \dot{\sigma}}\right] }{\sigma ^{2}}-\frac{\omega _{0}\Gamma /2}{%
\omega _{0}^{2}+\Gamma ^{2}/4}\frac{\left[ \mathbf{\sigma \times }\left[
\mathbf{\sigma \times \dot{\sigma}}\right] \right] }{\sigma ^{3}}.
\label{omega0viasigdot}
\end{equation}
Here the scalar $\omega _{0}$ can be found from Eqs. (\ref{sigmaiEquil}) and
(\ref{sigma0viasigma}):
\begin{eqnarray}
\omega _{0} &=&\frac{2k_{\mathrm{B}}T}{\hbar }\,\mathrm{arctanh}\left(
\sigma _{0}\right)  \nonumber \\
&\cong &\frac{2k_{\mathrm{B}}T}{\hbar }\,\mathrm{arctanh}\left( \sigma +%
\frac{1}{\Gamma }\frac{\left( \mathbf{\sigma \cdot \dot{\sigma}}\right) }{%
\sigma }\right)  \nonumber \\
&\cong &\frac{2k_{\mathrm{B}}T}{\hbar }\,\mathrm{arctanh}\left( \sigma
\right) +\frac{2k_{\mathrm{B}}T}{\hbar \Gamma }\frac{\dot{\sigma}}{1-\sigma
^{2}}.  \label{omega0scalarviasigmadot}
\end{eqnarray}
The $\dot{\sigma}$ correction here should be taken into account in the first
term of Eq. (\ref{omega0viasigdot}).\

\subsection{Linear mobility of domain walls}

The time derivative of the magnetic energy of the sample per unit
cross-sectional area due to dissipation has the form
\begin{equation}
\dot{U}=-\int_{-\infty }^{\infty }dz\,\hbar \mathbf{\omega }_{0}(z)\cdot
\mathbf{\dot{\sigma}}(z).  \label{Udot}
\end{equation}
On the other hand, motion of the DW results in the change of the energy of
the spins in the external field at the rate $-\sigma _{\infty }W_{\mathrm{ext%
}}v_{\mathrm{DW}},$ where $\sigma _{\infty }>0$ is the spin polarization in
domains and $W_{\mathrm{ext}}$ is given by Eq. (\ref{WiDef}). Equating this
rate to Eq. (\ref{Udot}), one obtains the energy balance relation
\begin{equation}
\sigma _{\infty }W_{\mathrm{ext}}v_{\mathrm{DW}}=\int_{-\infty }^{\infty
}dz\,\hbar \mathbf{\omega }_{0}(z)\cdot \mathbf{\dot{\sigma}}(z).
\label{EnergyBalance}
\end{equation}
This relation allows one to obtain the linear mobility of a domain wall as
an integral over the static DW profile without solving a complicated problem
of dynamical corrections to this profile. To find the linear mobility, one
has to express $\mathbf{\omega }_{0}(z)$ through $\mathbf{\dot{\sigma}}(z)$
using Eq. (\ref{omega0viasigdot}) that yields
\begin{equation}
\mathbf{\omega }_{0}\cdot \mathbf{\dot{\sigma}}=\omega _{0}\frac{\mathbf{%
\sigma }\cdot \mathbf{\dot{\sigma}}}{\sigma }+\frac{\omega _{0}\Gamma /2}{%
\omega _{0}^{2}+\Gamma ^{2}/4}\frac{\sigma ^{2}\mathbf{\dot{\sigma}}^{2}%
\mathbf{-}\left( \mathbf{\sigma \cdot \dot{\sigma}}\right) ^{2}}{\sigma ^{3}}%
.
\end{equation}
After that one has to use the fact that for a domain wall stationarily
moving in the positive $z$ direction, all quantities depend on the combined
space-like argument $\xi =z-v_{\mathrm{DW}}t,$ so that
\begin{equation}
\mathbf{\dot{\sigma}=-}v_{\mathrm{DW}}d\mathbf{\sigma }/dz.
\label{TimeDerviaxiDer}
\end{equation}
Using Eq. (\ref{omega0scalarviasigmadot}), one obtains the energy balance
equation in the form
\begin{eqnarray}
&&\sigma _{\infty }W_{\mathrm{ext}}  \nonumber \\
&=&\int_{-\infty }^{\infty }dz\,\left\{ \left( 2k_{\mathrm{B}}T\,\mathrm{%
arctanh}\left( \sigma \right) +\frac{2k_{\mathrm{B}}T}{\Gamma }\frac{v_{%
\mathrm{DW}}}{1-\sigma ^{2}}\frac{d\sigma }{dz}\right) \right.  \nonumber \\
&&\times \left( \frac{\mathbf{\sigma }}{\sigma }\cdot \frac{d\mathbf{\sigma }%
}{dz}\right) +\frac{\hbar v_{\mathrm{DW}}\lambda }{\,\mathrm{arctanh}\left(
\sigma \right) \,}\left[ 1+\frac{\lambda ^{2}\,/4}{\mathrm{arctanh}%
^{2}\left( \sigma \right) }\right] ^{-1}  \nonumber \\
&&\times \left. \left[ \frac{1}{\sigma }\left( \frac{d\mathbf{\sigma }}{dz}%
\right) ^{2}-\frac{1}{\sigma ^{3}}\left( \mathbf{\sigma }\cdot \frac{d%
\mathbf{\sigma }}{dz}\right) ^{2}\right] \right\} ,
\end{eqnarray}
where
\begin{equation}
\lambda \equiv \frac{\hbar \Gamma }{2k_{\mathrm{B}}T}.  \label{lambdaDef}
\end{equation}
Here the first term in the rhs can be easily integrated and gives a zero
contribution. Thus the rhs of Eq. (\ref{EnergyBalance}) becomes proportional
to $v_{\mathrm{DW}}^{2}.$ This is very fortunate since now one does not have
to take into account deviations from the static DW profile. In all practical
cases the inequality $\lambda \ll 1$ is strongly satisfied, thus one can
replace $\left[ \ldots \right] ^{-1}\Rightarrow 1.$ After that for the DW
speed in the linear regime one obtains
\begin{equation}
v_{\mathrm{DW}}=\frac{\sigma _{\infty }W_{\mathrm{ext}}}{2k_{\mathrm{B}}T}%
v_{\ast }=\frac{S\sigma _{\infty }g\mu _{\mathrm{B}}B_{z}}{k_{\mathrm{B}}T}%
v_{\ast },  \label{vDWFinal}
\end{equation}
where $v_{\ast }$ is the characteristic speed defined by
\begin{eqnarray}
v_{\ast }^{-1} &=&\int_{-\infty }^{\infty }dz\frac{1}{\Gamma }\left\{ \frac{1%
}{1-\sigma ^{2}}\left( \frac{d\sigma }{dz}\right) ^{2}\right.  \nonumber \\
&&+\left. \frac{\lambda ^{2}/2}{\sigma \,\mathrm{arctanh}\left( \sigma
\right) \,}\left[ \left( \frac{d\mathbf{\sigma }}{dz}\right) ^{2}-\left(
\frac{d\sigma }{dz}\right) ^{2}\right] \right\} ,  \label{vstarDef}
\end{eqnarray}
that resembles the expression for ferromagnets. \cite
{grakoe89,gar91llb,gar91edw,koegarharjah93prl,harkoegar95prb} Note that $%
\Gamma $ is not a constant and it has to be kept in the integrand. One can
see that if $\sigma \rightarrow \mathrm{1}$ in the domain wall, as is
usually supposed to be the case in ferromagnets, then $d\sigma
/dz\rightarrow 0,$ only the second term makes a contribution, and $\mu _{%
\mathrm{DW}}\varpropto 1/\Gamma .$ However, if $\sigma $ only slightly
deviates from 1, then because of the very small $\lambda $ the second term
becomes irrelevant and one obtains $\mu _{\mathrm{DW}}\varpropto \Gamma $
from the first term. For the linear (Ising-like) domain wall, $\mathbf{%
\sigma }=\sigma _{z}\mathbf{e}_{z},$ the expression in the square brackets
in second term of Eq. (\ref{vstarDef}) disappears. This is the case for Mn$%
_{12}$ except for very large transverse fields and very low temperatures.

The crossover between the two different types of the mobility behavior
occurs in the low-temperature range $T\ll \Delta /k_{\mathrm{B}},$ where the
deviation of $\sigma $ from 1 is small and given by Eq. (\ref{MagDefsig}).
Here in Eq. (\ref{vstarDef}) one can simplify
\begin{eqnarray}
1-\sigma ^{2} &\cong &2\left( 1-\sigma \right) \cong \frac{4k_{\mathrm{B}}T}{%
\Delta }\sqrt{1-\sigma _{z}^{2}}  \nonumber \\
\left( \frac{d\sigma }{dz}\right) ^{2} &\cong &\left( \frac{4k_{\mathrm{B}}T%
}{\Delta }\right) ^{2}\frac{\sigma _{z}^{2}}{1-\sigma _{z}^{2}}\left( \frac{%
d\sigma _{z}}{dz}\right) ^{2}  \nonumber \\
\left( \frac{d\mathbf{\sigma }}{dz}\right) ^{2} &=&\left( \frac{d\sigma _{z}%
}{dz}\right) ^{2}+\left( \frac{d\sigma _{x}}{dz}\right) ^{2}\cong \left(
\frac{d\sigma _{z}}{dz}\right) ^{2}  \nonumber \\
\mathrm{arctanh}\left( \sigma \right) &\cong &\frac{1}{2}\ln \frac{2}{%
1-\sigma }  \nonumber \\
&\cong &\frac{1}{2}\ln \left( \frac{\Delta }{2k_{\mathrm{B}}T}\left(
1-\sigma _{z}^{2}\right) ^{-1/2}\right) .
\end{eqnarray}
This yields
\begin{eqnarray}
v_{\ast }^{-1} &=&\int_{-\infty }^{\infty }dz\frac{1}{\Gamma }\left\{ \frac{%
4k_{\mathrm{B}}T}{\Delta }\frac{\sigma _{z}^{2}}{\left( 1-\sigma
_{z}^{2}\right) ^{3/2}}\right.  \nonumber \\
&&+\left. \frac{\lambda ^{2}}{\ln \left( \frac{\Delta }{2k_{\mathrm{B}}T}%
\left( 1-\sigma _{z}^{2}\right) ^{-1/2}\right) }\right\} \left( \frac{%
d\sigma _{z}}{dz}\right) ^{2}.
\end{eqnarray}
One can see that the second term dominates and the DW mobility has the form $%
\mu _{\mathrm{DW}}\varpropto 1/\Gamma $ only at extremely low temperatures,
\begin{equation}
T\lesssim \frac{\Delta }{k_{\mathrm{B}}}\left( \frac{\hbar \Gamma }{\Delta }%
\right) ^{2/3},  \label{FerroApplicability}
\end{equation}
up to the log term. In most region one has $\mu _{\mathrm{DW}}\varpropto
\Gamma .$

For Mn$_{12}$ in the typical case $\Delta \ll E_{\mathrm{D}}$ the DW is
Ising-like, and one should use the first term of Eq. (\ref{vstarDef}), where
$\sigma _{x}$ and $\sigma _{y}$ are negligibly small, so that in Eq. (\ref
{vDWFinal}) one has
\begin{mathletters}
\begin{equation}
v_{\ast }^{-1}=\int_{-\infty }^{\infty }dz\frac{1}{\Gamma }\frac{1}{1-\sigma
_{z}^{2}}\left( \frac{d\sigma _{z}}{dz}\right) ^{2}.  \label{vStarMn12}
\end{equation}
In the case of the unspecified relaxation rate $\Gamma $ that is considered
as a constant instead of Eq. (\ref{vDWFinal}) one obtains
\end{mathletters}
\begin{equation}
v_{\mathrm{DW}}=\frac{\sigma _{\infty }W_{\mathrm{ext}}}{k_{\mathrm{B}}T}%
\Gamma l_{\ast }
\end{equation}
with
\begin{mathletters}
\begin{equation}
l_{\ast }^{-1}=\int_{-\infty }^{\infty }dz\frac{1}{1-\sigma _{z}^{2}}\left(
\frac{d\sigma _{z}}{dz}\right) ^{2}.  \label{lstarDef1}
\end{equation}
At $T\ll T_{\mathrm{C}}$ according to Eq. (\ref{sigmazEqT0}) one has $%
l_{\ast }\sim R$ independent of $E_{\mathrm{D}}$ with a coefficient
dependent only on the lattice structure. Thus obtains $\mu _{\mathrm{DW}%
}\varpropto 1/T$ at low temperatures. In this case numerical calculation
yields $\mu _{\mathrm{DW}}$ diverging both at $T\rightarrow 0$ and $%
T\rightarrow T_{\mathrm{C}}$ and having a minimum at about 0.25$T_{\mathrm{C}%
}.$ One can improve the numerical procedure eliminating division 0/0 in
domains at $T\rightarrow 0$ using Eq. (\ref{DWProfileEqHT}) and
\end{mathletters}
\begin{equation}
\frac{1}{1-\sigma _{z}^{2}}\frac{d\sigma _{z}}{dz}=\frac{d}{dz}\mathrm{%
arctanh}\left( \sigma _{z}\right) .
\end{equation}
This yields
\begin{equation}
l_{\ast }^{-1}=\frac{E_{\mathrm{D}}}{k_{\mathrm{B}}T}\int_{-\infty }^{\infty
}dz\frac{d\sigma _{z}}{dz}\frac{dD_{zz}}{dz}.  \label{lstarDzz}
\end{equation}
This expression is very convenient since at low temperatures $D_{zz}\sim k_{%
\mathrm{B}}T/E_{\mathrm{D}}$ inside the DW that makes no computational
problems. The non-monotonic dependence of $v_{\mathrm{DW}}$ on temperature
should be taken with caution, however, because it is related to our
assumption that $\Gamma $ is a constant.

For $\Gamma $ given by Eq. (\ref{Gammai}) is intimately interwoven with the
rest of the mobility formula. Using Eq. (\ref{DWProfileEq}) in the form
\begin{equation}
\sigma _{z}=\frac{W(z)}{\hbar \omega _{0}}\tanh \frac{\hbar \omega _{0}}{2k_{%
\mathrm{B}}T},  \label{DWProfileEq1}
\end{equation}
one can transform Eq. (\ref{Gammai}) into the form
\begin{equation}
\Gamma =\frac{S^{2}\Delta ^{2}\left( g\mu _{B}B_{\bot }\right) ^{2}}{12\pi
E_{t}^{4}}\frac{W(z)}{\hbar \sigma _{z}(z)}.
\end{equation}
eliminating $W(z)$ with the use of Eq. (\ref{DWProfileEqHT}) one obtains
\begin{equation}
\Gamma =\frac{S^{2}\Delta ^{2}\left( g\mu _{B}B_{\bot }\right) ^{2}}{12\pi
E_{t}^{4}}\frac{2k_{\mathrm{B}}T}{\hbar }\frac{\mathrm{arctanh}\left( \sigma
_{z}\right) }{\sigma _{z}}.  \label{GammaArcTanh}
\end{equation}
Now Eq. (\ref{vDWFinal}) can be rewritten as
\begin{equation}
v_{\mathrm{DW}}=2\sigma _{\infty }\Gamma _{0}^{(\mathrm{ext})}l_{\ast },
\label{vDWvialstar}
\end{equation}
where
\begin{equation}
\Gamma _{0}^{(\mathrm{ext})}\equiv \frac{S^{2}\Delta ^{2}\left( W_{\mathrm{%
ext}}/\hbar \right) \left( g\mu _{B}B_{\bot }\right) ^{2}}{12\pi E_{t}^{4}}
\label{GammaExtDef}
\end{equation}
is the prefactor in Eq. (\ref{Gammai}) with $\hbar \omega _{0}\Rightarrow W_{%
\mathrm{ext}}$ and $l_{\ast }$ is a characteristic length defined by
\begin{mathletters}
\begin{equation}
l_{\ast }^{-1}=\int_{-\infty }^{\infty }dz\frac{\sigma _{z}}{\mathrm{arctanh}%
\left( \sigma _{z}\right) }\frac{1}{1-\sigma _{z}^{2}}\left( \frac{d\sigma
_{z}}{dz}\right) ^{2}.  \label{lstarDef}
\end{equation}
Similarly to Eq. (\ref{vStarMn12}) with $\Gamma \rightarrow 1,$ this
integral is finite at $T\rightarrow 0.$ To improve its numerical convergence
at low temperatures, one can rewrite it as
\end{mathletters}
\begin{mathletters}
\begin{eqnarray}
l_{\ast }^{-1} &=&\int_{-\infty }^{\infty }dz\frac{\sigma _{z}}{\mathrm{%
arctanh}\left( \sigma _{z}\right) }\frac{d\sigma _{z}}{dz}\frac{d\mathrm{%
arctanh}\left( \sigma _{z}\right) }{dz}  \nonumber \\
&=&\int_{-\infty }^{\infty }dz\frac{\sigma _{z}}{D_{zz}}\frac{d\sigma _{z}}{%
dz}\frac{dD_{zz}}{dz}.  \label{lstarDef2}
\end{eqnarray}
Again, at low temperatures $l_{\ast }\sim R$ with a coefficient depending
only on the lattice structure, c.f. Eq. (\ref{lstarDzz}).

Numerical results for temperature dependence of the velocity for $\Gamma =%
\mathrm{const}$ and for $\Gamma $ given by a direct phonon process in a
transverse magnetic field \cite{chugarsch05prb} are shown in Fig. \ref
{Fig-mu}.

\begin{figure}[tbp]
\includegraphics[angle=-90,width=8cm]{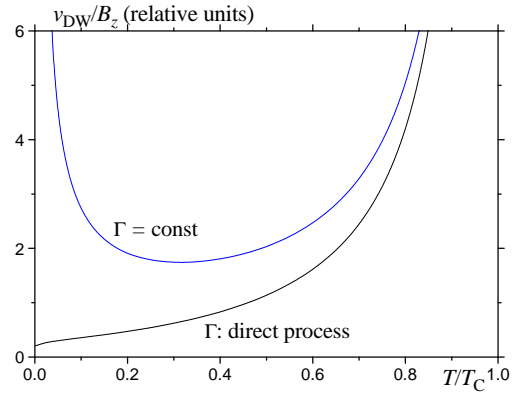}
\caption{{}Temperature dependences of the DW mobility in the cases of $%
\Gamma =\mathrm{const}$ and of $\Gamma $ due to the direct phonon processes.}
\label{Fig-mu}
\end{figure}

More generally, one can consider the relaxation rate of the form
\end{mathletters}
\begin{equation}
\Gamma =A\left( \hbar \omega _{0}\right) ^{n}\coth \frac{\hbar \omega _{0}}{%
2k_{\mathrm{B}}T},  \label{GammaGeneral}
\end{equation}
where $A=\mathrm{const}$ at low temperatures. In the case $\Delta \ll E_{%
\mathrm{D}}$ \ one has $\hbar \omega _{0}\cong W(z),$ then with the help of
Eq. (\ref{DWProfileEqHT}) one can rewrite $\Gamma $ as
\begin{equation}
\Gamma =A\left( 2k_{\mathrm{B}}T\,\mathrm{arctanh}\left( \sigma _{z}\right)
\right) ^{n}\frac{1}{\sigma _{z}}.
\end{equation}

\section{Discussion}

We have demonstrated that elongated crystal of magnetic dipoles arranged in
a body centered tetragonal lattice should exhibit ferromagnetic ordering.
For Mn$_{12}$ Acetate the corresponding Curie temperature computed within
mean-field model is about 0.8 K, which is close to the ordering temperature
0.9 K reported in recent neutron scattering experiments.\cite{luisetal05prl}
Other magnetic phases are separated from the ferromagnetic phase by a large
energy gap, thus making ferromagnetism in Mn$_{12}$ not difficult to
observe. This suggests that the analysis of all previous data on Mn$_{12}$
Acetate that were taken below 0.8 K should be re-examined for possible
effects of ferromagnetic order.

Dipolar-ordered crystals of molecular magnets should possess domain walls.
In a long crystal of length $L$ and radius $R\ll L$, the typical width of
the domain wall is of order $R$. When a small bias magnetic field $B_{z}$ is
applied, the domain wall moves at a speed $v_{\mathrm{DW}}\sim \lbrack {%
Sg\mu _{\mathrm{B}}B_{z}}/{(k_{\mathrm{B}}T)}]\langle \Gamma \rangle R,$
where $\langle \Gamma \rangle $ is the average spin relaxation rate. At,
e.g., $S=10$, $B_{z}=0.1$ T, $T=1$ K, and $R=1$ mm, this gives $v_{\mathrm{DW%
}}\sim 1$ m/s for $\langle \Gamma \rangle =10^{3}$ s$^{-1}$ and $v_{\mathrm{%
DW}}\sim 10^{3}$ m/s for $\langle \Gamma \rangle =10^{6}$ s$^{-1}$. It
should be emphasized that contrary to superradiance and laser effects in
molecular magnets, \cite{chugar02prl,chugar04prl} quantum dynamics of domain
walls is robust with respect to inhomogeneous broadening of spin levels and
phase decoherence of spin states. Crossing of spin levels due to a moving
front of dipolar field is sufficient for the effect to exist. Neither very
narrow spin levels nor phase coherence of spins in the domain wall are
required. We, therefore, believe that this effect should not be difficult to
observe in molecular magnets.

Our conclusion that low-temperature ferromagnetic phase is unstable against
division into domains may seem to contradict experiments on Mn$_{12}$ in
which, at low temperature, the crystal was shown to maintain finite
magnetization in a zero field for a very long time. This apparent
contradiction becomes resolved if one notices that in the zero-temperature
limit any evolution of the magnetic state of the Mn$_{12}$ crystal can only
occur through quantum tunneling between spin levels. The measure of this
tunneling is the splitting $\Delta $ which for low lying spin-levels is
negligibly small in zero field. To increase $\Delta $ one should apply
transverse magnetic field. Note that some evidence that transverse field is
needed to achieve local ferromagnetic order was obtained in Ref. %
\onlinecite{luisetal05prl}. In the absence of the longitudinal field, the
existence of the global ferromagnetic order would be hindered by the
presence of domain walls. One would need local measurements of the sample to
directly confirm ferromagnetic order below the Curie temperature. However,
in the presence of sufficient transverse field, a relatively weak
longitudinal field of the order of the dipolar field should be able to drive
the domain walls out, thus resulting in the uniform magnetization. Another
simple experiment can be conducted to observe the domain wall entering the
crystal. First the crystal should be uniformly magnetized by a strong
longitudinal field. Than the longitudinal field should be switched off and a
sufficient transverse field should be turned on. Finally, the compensating
longitudinal field should be applied. As soon as the resonant condition is
achieved, the domain wall should enter the crystal, resulting in the drastic
change of the magnetization. Some of these effects must have been already
observed in Mn$_{12}$ crystals but attributed to resonant quantum flipping
of magnetic molecules at random sites, instead of coherent motion of domain
walls.

The motion of a domain wall should not be confused with magnetic
deflagration. \cite{suzetal05prl,heretal05prl, garchu07prb} The latter is
also characterized by a moving front that separates regions with opposite
magnetization. However, while deflagration is driven by the release of heat
due to a strong field bias, the motion of a domain wall is driven by
Landau-Zener transitions in a relatively small bias. In experiment, the two
effects can be easily distinguished from each other if one notices that the
mobility of the domain wall depends strongly on the transverse field while
deflagration must have little dependence on the transverse field.

In the Introduction it was pointed out that domain walls in elongated
crystals of molecular magnets are propagating waves of Landau-Zener
transitions. The tunneling and relaxational dynamics of this process is
described by the density-matrix equation (\ref{DME-GGResAdi0}) that has a
form of the equation of motion for the magnetization, Eq. (\ref
{DME-sigmaEqVec}) in the two-state $\left| \pm S\right\rangle $
approximation. The energy sweep in the LZ effect is due to the change of the
dipolar field created by quantum spin transitions $\left| S\right\rangle
\rightleftharpoons \left| -S\right\rangle $ in the moving front. Thus, in
contrast to the standard LZ effect with externally controlled time-linear
sweep, the LZ effect considred here includes a \emph{self-consistent} sweep
that turns out to be time nonlinear. Indeed, the spatial profile of the
dipolar field in the DW in Fig. \ref{Fig-DW-Dzz_profile_Temperatures} is
strongly nonlinear at low temperatures. As the domain wall moves, a
time-nonlinear sweep is generated. LZ effect with time-nonlinear sweeps
generated by interaction between two-level systems have been studied in Ref. %
\onlinecite{gar03prb}. It was shown that if the time dependence $W(t)$
becomes steeper near the resonance, the staying probability $P$ increases,
whereas in the opposite case the transition probability $1-P$ increases.
Sweeps that become flat near the resonance lead to a nearly complete
transition, see, e.g., Fig. 7 of Ref. \onlinecite{gar03prb}. This is also
the case for the sweep corresponding to the $T\rightarrow 0$ curve in Fig.
\ref{Fig-DW-Dzz_profile_Temperatures}. In this case the LZ transition is
complete, $P=0,$ that is exactly what should be for a moving DW. Externally
controlled sweeps leading to a complete transition have been studied in Ref. %
\onlinecite{garsch02epl}. The case of moving domain walls is another
realization of the full-transition Landau-Zener effect.

Finally we would like to comment on the square root of time magnetic
relaxation observed in Mn-12 at low temperature. Existing theoretical works
attempted to explain this behavior by flips of molecular spins at random
sites of the crystal. \cite{alofer01prl,feralo05prb,cucetal99epjb} They
missed the fact that the low-temperature phase of Mn$_{12}$ Acetate is
ferromagnetic. Relaxation in the ferromagnetic phase must be determined by
the motion of domain walls. Within such picture the square root relaxation
may have simple explanation through random walk of the domain wall. If the
wall moves randomly at a constant rate to the right or to the left, the
resulting displacement and, thus, the change in the magnetization should be
proportional to the square root of the number of steps, that is, to the
square root of time. Such random walk may be thermal or it may be caused by
the adjustment of the domain wall to local solvent disorder or to local
hyperfine fields.

\section*{Acknowledgements}

D. G. would like to acknowledge stimulating discussions with Andrew Kent and
Gr\'{e}goire de Loubens, as well as to thank Kyungwha Park and Chris Beedle
for the information on the crystal structure of Mn$_{12}$ Acetate and
Wolfgang Wernsdorfer for the information on the crystal structure of Fe$_{8}$%
. This work has been supported by the NSF Grant No. DMR-0703639.

\appendix

\section{Calculation of the dipolar field in a crystal lattice}

The dipolar field $\mathbf{B}_{i}^{(D)}$ created on the lattice site $i$ by
all other spins, $j\neq i,$ aligned along the $z$ axis can be written in the
form
\begin{equation}
\mathbf{B}_{i}^{(D)}=\frac{Sg\mu _{B}}{v_{0}}\mathbf{D}_{i,z}
\label{HzMDzzRelation}
\end{equation}
with the dimensionless vector $\mathbf{D}_{i,z}$ given by
\begin{equation}
\mathbf{D}_{i,z}\equiv \sum_{j}\mathbf{\phi }_{ij}\sigma _{jz}.
\label{DzGenDef}
\end{equation}
Here $\mathbf{\sigma }$ is the Pauli matrix and
\begin{equation}
\mathbf{\phi }_{ij}=v_{0}\frac{3\mathbf{n}_{ij}(\mathbf{e}_{z}\cdot \mathbf{n%
}_{ij})-\mathbf{e}_{z}}{r_{ij}^{3}},\qquad \mathbf{n}_{ij}\equiv \frac{%
\mathbf{r}_{ij}}{r_{ij}}  \label{phivecijDef}
\end{equation}
with $v_{0}$ being the unit-cell volume. As said above, $\mathbf{D}_{i,z}$
of Eq. (\ref{HzMDzzRelation}) can be represented as a sum of the
contributions from the molecules inside and outside the sphere $r_{0}$
around the site $i$ satisfying $v_{0}^{1/3}\ll r_{0}\ll L,$ where $L$ is the
(macrocopic) linear size of the sample. The field from the spins at sites $j$
inside this sphere $r_{0}$ can be calculated by direct summation over the
lattice, whereas the field from the spins outside the sphere can be obtained
by integration. Replacing the index $i$ by the argument $\mathbf{r,}$ one
can write
\begin{equation}
\mathbf{D}_{z}(\mathbf{r})=\mathbf{D}_{z}^{(\mathrm{sph})}(\mathbf{r})+%
\mathbf{D}_{z}^{\prime }(\mathbf{r}),  \label{Dzr}
\end{equation}
where
\begin{equation}
\mathbf{D}_{z}^{(\mathrm{sph})}(\mathbf{r})\cong \bar{D}_{z}^{(\mathrm{sph}%
)}\sigma _{z}(\mathbf{r})\mathbf{e}_{z}  \label{DsphDef}
\end{equation}
and
\begin{equation}
\mathbf{D}_{z}^{\prime }(\mathbf{r})=\frac{\nu }{v_{0}}\int_{\left| \mathbf{%
r-r}^{\prime }\right| >r_{0}}d^{3}r^{\prime }\mathbf{\phi }\left( \mathbf{r-r%
}^{\prime }\right) \sigma _{z}(\mathbf{r}^{\prime }).  \label{DzIntegral}
\end{equation}
In Eq. (\ref{DsphDef}) it is assumed that the change of $\sigma _{z}(\mathbf{%
r})$ inside the sphere $r_{0}$ is negligible. $\bar{D}_{z}^{(\mathrm{sph})}$
is a constant depending on the lattice structure. In Eq. (\ref{DzIntegral}) $%
\nu $ is the number of molecules per unit cell,
\begin{equation}
\mathbf{\phi }\left( \mathbf{r}\right) =v_{0}\frac{3\mathbf{n}_{\mathbf{r}}(%
\mathbf{e}_{z}\cdot \mathbf{n}_{\mathbf{r}})-\mathbf{e}_{z}}{r^{3}}=v_{0}%
\mathrm{rot}\frac{\left[ \mathbf{e}_{z}\times \mathbf{n}_{\mathbf{r}}\right]
}{r^{2}},  \label{phivecDef}
\end{equation}
and $\mathbf{n}_{\mathbf{r}}\equiv \mathbf{r/}r$. In the main part of the
paper one needs only the $z$ component of the vector $\mathbf{D}_{z},$ i.e.,
$D_{zz}.$

Using the integral formula
\begin{equation}
\int_{V}dV\,\mathrm{rot}\mathbf{F=}\int_{S}d\mathbf{S}\times \mathbf{F}
\label{rotFormula}
\end{equation}
and the relation
\begin{equation}
\mathrm{rot}\left[ f\mathbf{A}\right] =\left[ \nabla f\times \mathbf{A}%
\right] +f\mathrm{rot}\mathbf{A,}
\end{equation}
one can rewrite Eq.\ (\ref{DzIntegral}) as
\begin{eqnarray}
\mathbf{D}_{z}^{\prime }(\mathbf{r}) &=&\int_{S}d\mathbf{S}^{\prime }\times
\sigma _{z}(\mathbf{r}^{\prime })\frac{\left[ \mathbf{e}_{z}\times \left(
\mathbf{r^{\prime }-r}\right) \right] }{\left| \mathbf{r^{\prime }-r}\right|
^{3}}  \nonumber \\
&&-\int_{\left| \mathbf{r-r}^{\prime }\right| >r_{0}}d^{3}r^{\prime }\nabla
\sigma _{z}(\mathbf{r}^{\prime })\times \frac{\left[ \mathbf{e}_{z}\times
\left( \mathbf{r^{\prime }-r}\right) \right] }{\left| \mathbf{r^{\prime }-r}%
\right| ^{3}}  \nonumber \\
&&-\int_{S_{r_{0}}}d\mathbf{S}^{\prime }\times \sigma _{z}(\mathbf{r}%
^{\prime })\frac{\left[ \mathbf{e}_{z}\times \left( \mathbf{r^{\prime }-r}%
\right) \right] }{\left| \mathbf{r^{\prime }-r}\right| ^{3}}.
\label{DzprimeFront}
\end{eqnarray}
For a flat domain wall with
\begin{equation}
\nabla \sigma _{z}(\mathbf{r}^{\prime })=\frac{d\sigma _{z}(z^{\prime })}{%
dz^{\prime }}\mathbf{e}_{z}
\end{equation}
the second term in Eq. (\ref{DzprimeFront}) is zero. In this case one has
\begin{equation}
\mathbf{D}_{z}^{\prime }(\mathbf{r})=\nu \int_{S}d\mathbf{S}^{\prime }\times
\sigma _{z}(\mathbf{r}^{\prime })\frac{\left[ \mathbf{e}_{z}\times \left(
\mathbf{r^{\prime }-r}\right) \right] }{\left| \mathbf{r^{\prime }-r}\right|
^{3}}-\frac{8\pi \nu }{3}\sigma _{z}(\mathbf{r})\mathbf{e}_{z},
\label{DveczSavard}
\end{equation}
where the integral is taken over the surface of the sample, the vector $d%
\mathbf{S}^{\prime }$ is directed outwards, and the last term is the
integral over the sphere $r_{0}$, taken with $\sigma _{z}(\mathbf{r}^{\prime
})\Rightarrow \sigma _{z}(\mathbf{r}).$ The first term in Eq.\ (\ref
{DveczSavard}), multiplied by $g\mu _{B}S/v_{0},$ yields the \emph{%
macroscopic} internal field in the sample that also can be obtained from the
Biot-Savart formula with the molecular currents $\mathbf{j=}c\,\mathrm{rot\,}%
\mathbf{M}$ flowing on the sample's surface, $\mathbf{M=}\left( g\mu
_{B}S\nu /v_{0}\right) \mathbf{e}_{z}$ inside the sample and $\mathbf{M=0}$
outside the sample.

In particular, for $\sigma _{z}(\mathbf{r})=\sigma _{z}=\mathrm{const}$ for
ellipsoidal samples the integral in Eq. (\ref{DveczSavard}) becomes $4\pi
\nu \left( 1-n^{(z)}\right) \sigma _{z}\mathbf{e}_{z},$ where $n^{(z)}$ is
the demagnetizing factor$.$ This yields $\mathbf{D}_{z}^{\prime }(\mathbf{r}%
)=D_{zz}^{\prime }\sigma _{z}\mathbf{e}_{z},$ where $D_{zz}^{\prime }=4\pi
\nu \left( 1/3-n^{(z)}\right) $ . Then Eqs. (\ref{Dzr}) and (\ref{DsphDef})
result in Eq. (\ref{DzzEllipsoid1}) in the homogeneous case. For a cylinder
magnetized with $\sigma _{z}=$ $\sigma _{z}(z),$ the field along the
symmetry axis is given by Eq. (\ref{DzzCylinder}).

\bibliographystyle{apsrev}
\bibliography{gar-own,chu-own,gar-tunneling,gar-relaxation,gar-oldworks,gar-books,gar-general}

\end{document}